\documentclass{IET-Conf-Paper}

\usepackage{color}
\usepackage{threeparttable}
\usepackage{booktabs}
\usepackage{tabularx}
\usepackage{balance}
\usepackage{multirow,multicol, array}
\usepackage{algpseudocode}
\newcommand{\mytablefontsize}{9pt}
\newcommand{\mytablebaselineskip}{1.1}

\usepackage{mathtools}
\usepackage[linesnumbered,ruled]{algorithm2e}
\begin{document}

\title{A COGNITIVE FREQUENCY ALLOCATION STRATEGY FOR MULTI-CARRIER RADAR AGAINST COMMUNICATION INTERFERENCE}
\author{Zhao Shan, Lei Wang\corr\, Pengfei Liu, Tianyao Huang, Yimin Liu} 
\address{{Department of Electronic Engineering, Tsinghua University, Beijing, China}\\
\email{leiwangqh@tsinghua.edu.cn}}

\begin{abstract}
Modern radars often adopt multi-carrier waveform which has been widely discussed in the literature. However, with the development of civil communication, more and more spectrum resource has been occupied by communication networks. Thus, avoiding the interference from communication users is an important and challenging task for the application of multi-carrier radar. In this paper, a novel frequency allocation strategy based on the historical experiences is proposed, which is formulated as a Markov decision process (MDP). In a decision step, the multi-carrier radar needs to choose more than one frequencies, leading to a combinatorial action space. To address this challenge, we use a novel iteratively selecting technique which breaks a difficult decision task into several easy tasks. Moreover, an efficient deep reinforcement learning algorithm is adopted to handle the complicated spectrum dynamics. Numerical results show that our proposed method outperforms the existing ones.
\end{abstract}

\maketitle

\section{Introduction}
The multi-carrier radar, which adopts the waveform containing several carrier frequencies, has been widely discussed in the literature. The multi-carrier waveform brings multiple advantages to radar, e.g., the resistance to multipath interference~\cite{1631800}, the ability to transmit communication information~\cite{8683655} and the potential to synthesise a large bandwidth~\cite{8361480}.
With the rapid growth of communication demand, more and more communication devices occupy a large amount of spectrum resources. Therefore the multi-carrier radar sometimes shares spectrum resources with communication users~\cite{greco2018cognitive}. In order to decrease the communication interference, a cognitive multi-carrier radar (CMCR) needs to extract information from the past observation and select frequencies to reduce the interference in the next transition.

To model the problem of spectrum sharing, the shared spectrum is divided into subbands (a subband is also called a channel which is widely used in communication network) and operated by the communication users in time-slot fashion~\cite{greco2018cognitive}~\cite{sheng2020sensing}. 
In~\cite{9527128}, the states of the channels are assumed to be independent in different time slots, and a contextual multi-armed bandit is used to achieve an online learning framework. In~\cite{liu2010indexability}, a conventional model-based method assumes the state of each subband is independent and follows a 2-state Markov chain, where the whittle index policy achieves the optimal performance. However, in a real scenario, the states of the shared spectrum are usually not independent between different time slots and channels. 

Recently, deep reinforcement learning (DRL) is becoming a competitive method especially suitable for the problem of channels selection when the channel states are correlated thanks to the model-free property. In~\cite{wang2018deep}, The authors firstly tackle the problem of channel selection with the deep Q-learning algorithm but they only consider the case where only one channel is selected. In~\cite{zhong2019deep}, a deep actor-critic reinforcement learning algorithm is employed for multi-channel selection. However, the authors adopts
an action space with $\binom{K}{M}$ choices, where $K$ is the number of the channels and $M$ is the number of the frequencies of the CMCR. This leads to a combinatorial action space which is hard to learn especially when $M$ is large.

In this paper, we formulate the CMCR frequency selection problem as a Markov decision process and adopt the DRL framework to solve it. To tackle the challenge of combinatorial action space, we use a novel iteratively selecting technique~\cite{song2019solving} which converts a difficult decision task into several easy ones. Different from the Q-learning algorithm in~\cite{wang2018deep} and the actor-critic algorithm in~\cite{zhong2019deep}, we employ
the proximal policy optimization (PPO) algorithm~\cite{schulman2017proximal} to train our proposed DRL model, which greatly improves the learning efficiency. To demonstrate the benefits of cognition, we show that the detection probability of a target is improved greatly compared with a non-cognitive multi-carrier radar. Numerical simulations show that our proposed method has good robustness under different channel parameters.
 
The rest of the paper is organized as follows. Section $2$ formulates the system model and the decision problem. In Section $3$, we give a detailed description of our proposed method. Then the numerical results of our proposed method is shown in section $4$. Finally we give a brief summary of the paper in section $5$. 

\section{Problem Formulation}
In this section, we introduce the signal model of the CMCR, the model of the shared spectrum and the formulation of the frequency selection problem for the CMCR. 
\subsection{Signal Model}
\subsubsection{CMCR Signal}
The CMCR transmits a multi-band waveform containing several monotone signals. Denote the set of available carrier frequencies as $\mathcal{F} = \{f_c + k\Delta f|k \in \mathcal{K}\}$, where $f_c$ is the initial carrier frequency, $\mathcal{K}=\{0,1,...,K-1\}$, $K$ is the number of available frequencies, and $\Delta f$ is the frequency step. At the beginning of a time slot, the CMCR selects a set of carrier frequencies $\mathcal{B}$ from $\mathcal{F}$, where $\mathcal{B}=\{f_m=f_c+d_m\Delta f|d_m \in \mathcal{M}\}, \mathcal{M} \subset \mathcal{K}$. The cardinality of $\mathcal{M}$ is constant, i.e., $\lvert \mathcal{M} \rvert = M$. The waveform with frequency $f$ can be written as $\phi(f,t)=\text{rect}(t/T_p)e^{j2 \pi ft}$, where $T_p$ represents the pulse duration, and $\text{rect}(t)=1$ for $t\in[0,1)$ and zero otherwise. In each time slot, the CMCR transmits $N$ pulses and the $n$-th pulse at time instance $t$ is
\begin{equation}
\label{eq.sendingwave}
T(n,t)=\sum_{m=0}^{M-1}\frac{1}{\sqrt{M}}\phi(f_c+d_m\Delta f, t-nT_r),
\end{equation}
where $n=0,1...N-1$, $T_r$ is pulse repetition interval and the factor $\frac{1}{\sqrt{M}}$ guarantees that the transmitted waveform has the same total power under different $M$.

Assume that there exists a Swelling-0 target moving along the radar line of sight with velocity $v$ and the range between the target and the CMCR is $r(0)$ at the time instance $t=0$. Under the ``stop and go'' model~\cite{richards2014fundamentals}, it holds that $r(t) \approx r(0)+nvT_r$. After being reflected by the target, the signal propagates back to the radar antenna with delay $2r(t)/c$. The $n$-th echo is
\begin{equation}
\label{eq.rec_continum}
R(n,t)=\sum_{m=0}^{M-1}\frac{\beta}{\sqrt{M}}\phi(f_c+d_m\Delta f, t-nT_r-2\frac{r(0)+nvT_r}{c}),
\end{equation}
where $\beta \in \mathbb{C}$ is the scattering coefficient, $c$ is the speed of light. Then, the received echoes are down-converted to the baseband using carrier frequencies $f_m=f_c+d_m\Delta f$. Under the assumption that the transmitted waveforms with different frequencies are orthogonal with each other, the echoes with different frequencies can be separated in the receiver. At the frequency $f_m$, we sample the baseband echoes at the time instance $t=nT_r+iT_p, i=0,1...\lfloor T_r/T_p\rfloor - 1$, such that each pulse is sampled once. The sampled echoes at different $i$ is processed independently and individually. Assume that the target is present in the $i$-th range cell. Then the signal of the $n$-th pulse at the frequency $f_m$ can be written as (see \cite{9354050} for details)
\begin{equation}
\label{eq.receivingwave}
x_{m,n}=\frac{\beta}{\sqrt{M}}e^{-j4\pi f_mr/c}e^{-j4\pi f_mnvT_r/c}.
\end{equation}
\subsubsection{Communication Interference Signal}
Denote the receiver thermal noise of frequency $f_m$ as $w_{m,n}$ and $w_{m,n}\sim \mathcal{CN}(0, N_0\Delta f)$, where $N_0$ is average power spectral density and $\mathcal{CN}(\cdot)$ denotes the probability density function of complex Gaussian distribution. If the frequency $f_m$ is occupied by the communication users, the interference is described as $j_{m,n}\sim \mathcal{CN}(0, J_m\Delta f)$~\cite{wang2019network}, where $J_m$ is the interference average power spectral density of the frequency $f_m$. Let $i_{m,n}=w_{m,n}+1(d_m)j_{m,n}$, where $1(d_m)$ is an indicator to show whether the frequency $f_m$ is occupied by communication users. To summarize, the matrix form of the noisy received echoes is
\begin{equation}
\label{eq.receivingwaveMatrix}
    \textbf{Y}=\textbf{X}+\textbf{I},
\end{equation}
where $\textbf{Y},\textbf{X},\textbf{I}\in\textbf{C}^{M\times N}$, and $[\textbf{X}]_{m,n}=x_{m,n},[\textbf{I}]_{m,n}=i_{m,n}$. 
\subsubsection{Target Detection Method}
 To detect the target, the received echoes of the same frequency are integrated coherently using matched filtering. Since the CMCR may occupy several separated frequency bands, we apply noncoherent integration between different frequencies ~\cite{li2021robust}. Specifically, the concerned velocity range is divided into several grids, where we apply target detection identically and individually. For any velocity grid $v_t$, the coherent integration result of each frequency is collected as $\textbf{C}=[c_1,...,c_{M}]$, where $c_m=\sum_{n=0}^{N-1}y_{m,n}e^{j4\pi f_mnv_tTr/c}$. Then the test statistic is~\cite{kay1993fundamentals}
\begin{equation}
\label{test_statistic}
    \textit{T}(\textbf{C})=\sum_{m=0}^{M-1}\frac{2\lvert c_m\rvert^2}{N\Delta f(N_0+1(d_m)J_m)}.
\end{equation}
Under hypothesis $H_0$ (the target is absent), $\textbf{Y}=\textbf{I}$, $c_m\sim \mathcal{CN}(0, N\Delta f(N_0 + 1(d_m)j_{m,n}))$. $\textit{T}(\textbf{C})$ follows a Chi-square distribution of $2M$ degrees of freedom, and the false alarm rate is
\begin{equation}
\label{eq.false_rate}
    p_{\text{f}}=P_{\text{r}}(\textit{T}(\textbf{C})\leq T_{\text{th}}|H_0)=Q_{2M}(T_{\text{th}}),
\end{equation}
where $T_{\text{th}}$ is the decision threshold and $Q_{2M}(\cdot)$ is the cumulative distribution of the $2M$-degree Chi-square distribution.

Under hypothesis $H_1$ (a target with velocity $v_t$ is present), $\textbf{Y}=\textbf{X}+\textbf{I}$, after simple derivation, $c_m\sim \mathcal{CN}(\frac{N\beta_m}{\sqrt{M}}, N\Delta f(N_0 + 1(d_m)J_{m}))$, where $\beta_m=\beta e^{-j4\pi f_mr/c}$. As a result, $\textit{T}(\textbf{C})$ follows non-centralized $2M$-degree Chi-square distribution whose center is $\frac{N\lvert \beta \rvert^2}{M}\sum_{m=0}^{M-1}\frac{1}{\Delta f(N_0+1(d_m)J_m)}$. Therefore, the probability of detection is
\begin{equation}
\label{eq.detection_rate}
\begin{aligned}
    p_{\text{d}} &\!=\!P_{\text{r}}(\textit{T}(\textbf{C})>T_{\text{th}}|H_1)\\&\!=\!1\!-\!Q_{2M,\text{NC}}(T_{\text{th}}, \frac{N\lvert \beta \rvert^2}{M}\sum_{m=0}^{M-1}\frac{1}{\Delta f(N_0+1(d_m)J_m)}),
\end{aligned}
\end{equation}
where $Q_{2M,\text{NC}}(\cdot, \cdot)$ is the cumulative distribution of non-centralized 2$M$-degree Chi-square distribution. Given $p_\text{f}$, the decision threshold $T_{\text{th}}$ is calculated as (\ref{eq.false_rate}), then the detection rate $p_{\text{d}}$ can be obtained as (\ref{eq.detection_rate}).

\subsection{Spectrum Model}
The shared spectrum is divided into $K$ licensed channels, where each channel has the same bandwidth $\Delta f$, and the centers of the channels locate in $\mathcal{F}$. We consider each channel in the communication network has two possible states: \textbf{good}(1) which means there is no communication user accessing this channel or \textbf{bad}(0) which means the opposite. Therefore the switching pattern of the shared spectrum can be described as a $2^K$-state Markov chain, whose transition matrix is denoted by $\textbf{P}$. It is hard to describe a $2^K$-state Markov chain completely. Thus, several simplified models have been proposed, such as the fixed-pattern switching model and the correlated Markov model~\cite{wang2018deep}:
\begin{itemize}
\item In the fixed-pattern switching model, the channels are divided into several subsets, where the channels in a same subset have the same state. The states of these subsets become good with a switching probability $p_{\text{sw}}$ in a fixed order.
\item In the correlated Markov model, there are only two or three independent channels. The state of each independent channel follows a 2-state Markov chain with transition probabilities $\{p_{ij}\}_{i,j=0,1}$. Other channels are identical or opposite to one of these independent channels.
\end{itemize}

\begin{figure}[th]
    \centering
    \includegraphics[width=0.30\textwidth,trim={0pt 0pt 0pt 0pt},clip]{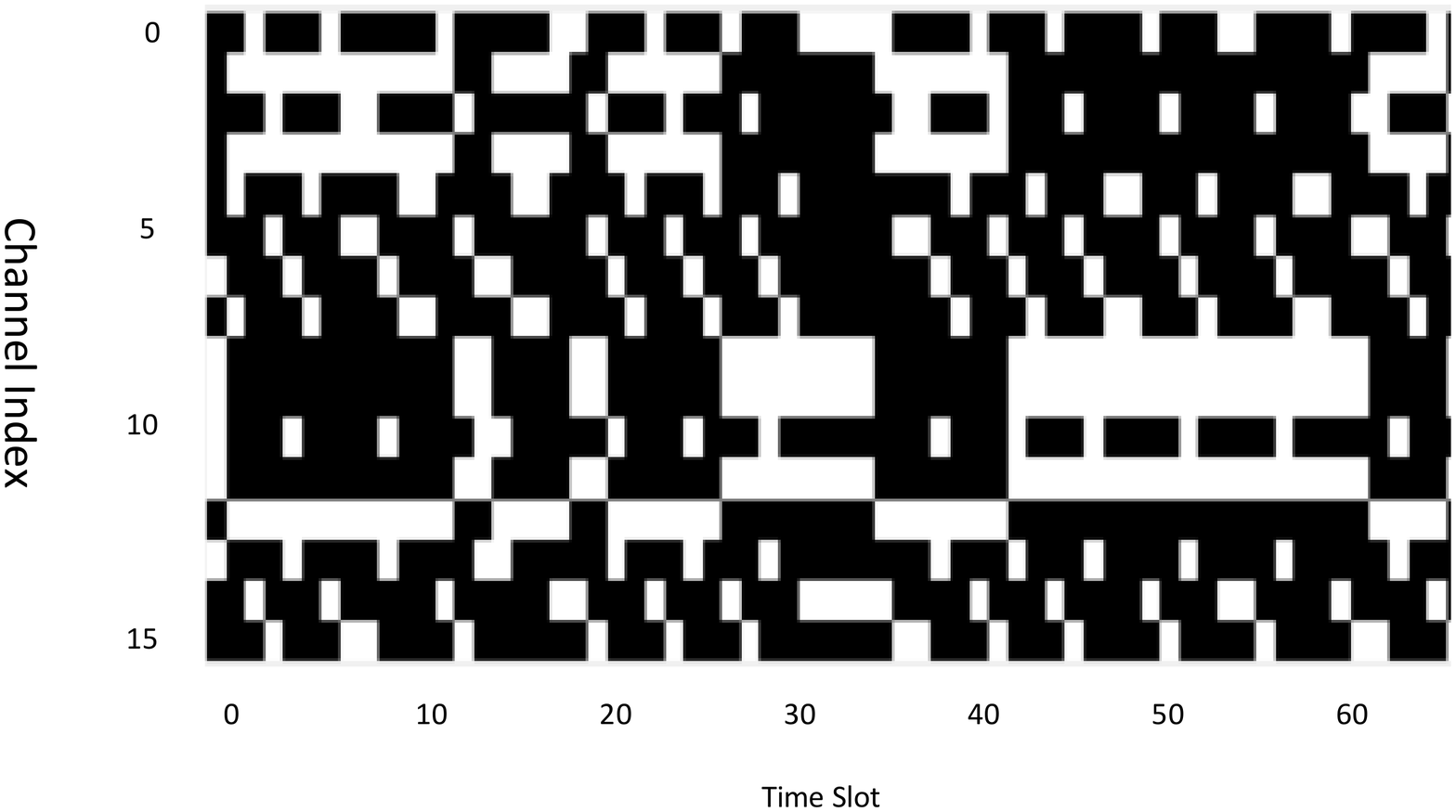}
    \vspace{-2em}
    \caption{An example of the channel evolution.}
    \label{fig:channe}
\end{figure}

In practice, the evolution of communication network is pretty complicated which is usually described using more than one spectrum models~\cite{xu2020application}. For example, we give a more realistic scenario where the communication network is a mixture of two models. All channels are divided into two subsets, one subset follows the fixed-pattern model and the other follows the correlated Markov model. In Fig.\ref{fig:channe}, we provide a pixel illustration to show how \textbf{good}/\textbf{bad} states change over $65$ time slots (white pixel means the channel is in good state in this time slot), where $10$ channels follow fixed-pattern switching model with $p_{\text{sw}}=0.8$ and $6$ channels follow correlated Markov model with $p_{01}=p_{10}=0.2$. It is important to point out that our proposed model is not limited to a particular spectrum model.
\vspace{-2em}
\vspace{-2em}

\subsection{Frequency selection problem}
We consider a detection process consisting of $L$ time slots. During the process, the CMCR radar aims to maximize the average detection rate through frequency selection. According to (\ref{eq.detection_rate}), given $p_{\text{f}}$, the detection probability of a target is related to the scattering intensity and the interference power of each selected frequency.
Therefore, the detection probability can be enhanced by minimizing the summation of the inverse of the interference power in the selected frequencies. The problem is formulated as
\begin{equation}\label{eq.our_problem}
\begin{aligned}
&\min_{\mathcal{M}_l,l=1,2...L} \quad \frac{1}{L} \sum_{l=1}^{L}\sum_{m=0}^{M-1}\frac{1}{1_l(d_m)J_m+N_0}\\
&\begin{array}{cc}
\text{s.t.} &  d_m\in\mathcal{M}_l\\
     &  \lvert\mathcal{M}_l\rvert=M.\\
\end{array}
\end{aligned}
\end{equation}
This is a difficult problem. Specifically, the states of the channels in the coming time slot are unknown to the CMCR, because it can only get $1_l(d_m)$ after the transmission in the $l$-th time slot, while the decision $\mathcal{M}_l$ has to be made at the beginning of $l$-th time slot. We need to give a frequency selection policy $\pi$ which is a mapping from the historical observation to a subset of $\mathcal{F}$. In each time slot, the CMCR chooses $M$ channels to transmit waveforms according to $\pi$. After the signal processing, the interference power of each selected channel is measured. Thus, the CMCR gets a partial observation of the state of spectrum. In the next time slot, the observation of the last time slot becomes historical information, based on which the CMCR makes a new decision. As a result, the historical actions will further influence the future action. Therefore, the aforementioned problem is a typical partially observable Markov decision process (POMDP) which can be solved by deep reinforcement learning especially for correlated channel environment~\cite{wang2018deep}.

\section{Methodology}
In the DRL framework, an agent interacts with an unknown environment. At each time step $t$, the agent observes a state $s_t$ of the environment and chooses an action $a_t$. Given $a_t$, the state of the environment evolves into a state $s_{t+1}$ following an unknown transition probability which depends on $s_t$ and $a_t$. At the same time, the agent gets a reward $r_{t+1}$ whose distribution also depends on $s_t$ and $a_t$. The agent aims to find a policy which maximizes the expected cumulative discounted reward: $\mathbf{E}(\sum_{t=0}^{\infty}\gamma^tr_t)$, where $\gamma\in(0,1]$ is a discount factor for the following rewards. In the DRL algorithm, an action is generated from the policy network (policy based algorithms) or the Q-value network (value based algorithms), we refer readers to~\cite{sutton1998introduction} for more details.

In the existing methods~\cite{zhong2019deep}, the state $s_t$ is encoded from the historical observation, which is fed into the policy network or Q-value network. The action $a_t$ is the working frequency set $\mathcal{M}$ generated  directly from the policy or value network. The reward $r_t$ is the summation of the interference power of the chosen channels. However, the number of valid actions is $\binom{K}{M}$, which leads to an exponential growth of the network output dimension. In this paper, we break a complete action into $M$ sub-actions, which means we only choose one channel in a decision step, and the consecutive $M$ decisions form a complete action. In other words, we break a general MDP into an equivalent but much easier MDP, namely iteratively selected MDP (ISMDP)~\cite{song2019solving}. 

\subsection{Observation, Action and Reward}
The action space, the observation space and the reward function of the proposed ISMDP is defined as follows:

The $i$-th sub-action in $l$-th time slot $a_{l,i}\in\{0,1,2...K-1\}, 0\le i < M, 1 \le l \le L$ is a discrete action. The $M$ sub-actions should not be repeated, which leads to $M-i$ choices in $a_{l,i}$.
The consecutive $M$ sub-actions form a complete action $a_l$. Compared with the method generating $a_l$ in one decision step~\cite{zhong2019deep}, every sub-decision is a much easier decision task.

The observation $o_{l, i}$ of ISMDP consists of the historical spectrum record $\textbf{O}_l$ and the mask vector recording the sub-actions that have been chosen before the $i$-th sub-decision. Similar to~\cite{zhong2019deep}, we keep the spectrum record of the most recent $P$ time slots. Thus, the spectrum record $\textbf{O}_l$ is a $K{\times}P$ matrix, where each column is the record of the spectrum information of a time slot. Since the CMCR can only obtain the states of $M$ chosen channels in one time slot, each column in $\textbf{O}_l$ consists of $M$ nonzero elements standing for the interference power of the selected channels. The positions of the unobserved channels are set to zero. At the end of a time slot, the newest record is added into $\textbf{O}_l$ and the record of $P$ slots before will be removed. The mask is a vector of length $M$, which is set to a zero vector before the $0$-th sub-action. If the $a_{l,i}$-th channel is selected in the $i$-th sub-action, the $a_{l,i}$-th element of the mask is set to one.

All of the sub-actions are made at the beginning of each time slot, while the radar can only get the interference power of each selected channel after the signal processing in this time slot. Therefore the reward of the proposed ISMDP is

\begin{equation}
\label{ppo.reward}
R_{l,i}=\left\{
\begin{aligned}
0 &, & i < M-1, \\
\sum_{m=0}^{M-1}1/(1_mJ_m+N_0) &, & i=M-1.
\end{aligned}
\right.
\end{equation}

\subsection{RL Algorithm and Network Design}
For convenience, we use $t=lM+i$ to stand for the time step of ISMDP. Thus, the length of the trajectory is $T=ML$.
To train the proposed reinforcement learning model, we adopt the PPO algorithm proposed in~\cite{schulman2017proximal} which contains a policy network and a value network. The PPO is a type of policy gradient algorithm. The policy network is used to generate actions based on the states and the value network is used to predict the expected value of the states which can reduce the training variance of the policy network. 
To further reduce the training variance, PPO uses a truncated version of generalized advantage estimation, which is calculated by 
\begin{equation}
    A_t = \delta_t + (\gamma\lambda)\delta_{t-1} +... +  (\gamma\lambda)^{T-t}\delta_{T-1}
\end{equation}
where $\delta_t = r_t + \gamma V(s_{t+1}) - V(s_t)$ is the temporal-difference error defined in \cite{sutton1998introduction} for the $t$-th step in a length-$T$ trajectory segment, $V(s_t)$ is the output of the value network, and $\lambda$ is the exponential weight for generalized advantage estimation introduced in~\cite{schulman2015high}.
Denote the parameters of policy network and the value network as $\theta$, then the surrogate loss function used in the policy network is 
\begin{equation}
\label{eqn.ppo_policy}
    L_t(\theta) = \min(R_t(\theta)A_t, \text{clip}(R_t(\theta), 1-\epsilon, 1+\epsilon)A_t),
\end{equation}
where $R_t(\theta)$ is the probability ratio defined in \cite{schulman2017proximal}, the clip function clips $R_t(\theta)$ to be inside the interval $[1-\epsilon, 1+\epsilon]$ and $\epsilon$ is a small value.
The surrogate loss for the value network is:
\begin{equation}
\label{eqn.ppo_value}
    L_t^{V} = (V_\theta(s_t) - V_t^{\text{targ}})^2,
\end{equation}
where $V_t^{\text{targ}}$ is the target cumulative reward of the state $s_t$ which is calculated using a bootstrapping method~\cite{sutton1998introduction}. 

The policy network and value network in the PPO agent are both multi-layer perceptron network consisting of $3$ hidden layers. The policy network takes the observation as input and outputs the action distribution. Since the action number varies between sub-actions, the output layer size of the policy network should also varies. Inspired by~\cite{song2019solving}, we share a same policy network between different sub-actions where the mask vector is used to zero out invalid choices. The value network takes observation as input and predicts the value of the current state. To speed up the convergence, the policy network and value network share the first two layers. 

We call our method iteratively selecting proximal policy optimization (ISPPO). The detailed training algorithm is shown in Algorithm \ref{alg:training}. The parameters of policy network and value network is learnt after $E$ iterations. In each iteration, the CMCR interacts with the shared spectrum for $N_{traj}$ times to collect training examples. In the $l$-th interaction, the DRL agent generates $M$ sub-actions iteratively forming a working frequency set $\mathcal{M}_l$. Then the CMCR transmits waveforms using the frequency set $\mathcal{M}_l$ and calculates the reward of the sub-actions as (\ref{ppo.reward}). After the sample collection, the PPO agent updates the parameters of the policy network and the value network according to (\ref{eqn.ppo_policy}) and (\ref{eqn.ppo_value}), respectively. 

\begin{algorithm}[tb]
\caption{Training Process of ISPPO}
\label{alg:training}
\KwIn{frequency number $M$\;
      Epoch number $E$\;
      Sampling trajectory number $N_{traj}$\; 
      Time slot number $L$;
}
\KwOut{The learned network parameters $\theta$;}
\For{$iteration \leftarrow 1$ \KwTo $E$}{
    Samples set $S$ = \{\};\\
    \For {$traj \leftarrow 1$ \KwTo $N_{traj}$}{
        $t=0$\;
        \For{$ l \leftarrow 1$ \KwTo $L$}{
            $\mathcal{M}_l = \{\}$\;
            Construct historical observation matrix $\textbf{O}_{l}$\;
            Initialize mask vector $mask=0^M$\;
            \For{sub action $i \leftarrow 0$ \KwTo $M-1$}{
                $o_{t}=\{\textbf{O}_{l}, mask\}$\;
                Sample a sub action $a_t$ from $\pi(a|o_t)$\; 
                $mask[a_t] = 1$\;
                $\mathcal{M}_l = \mathcal{M}_l \cup \{a_t\}$\;
                
                \If{$i=M-1$}{
                Transmit waveforms using $\mathcal{M}_l$ in slot $l$
                }
                Obtain $r_{t+1}$ as (\ref{ppo.reward})\;
                $S = S\cup(o_t, a_t, r_{t+1})$\;
                $t = t+1$;
                }
            }
        }
    Update the policy network as (\ref{eqn.ppo_policy})\;
    Update the value network as (\ref{eqn.ppo_value})\;
}
\Return $\theta$
\end{algorithm}

\section{Experiments and Numerical Results}
In this section, we validate the effectiveness of the proposed ISPPO method via simulation and compare it with three existing methods, i.e., the existing one step DRL~\cite{zhong2019deep}, the whittle index policy~\cite{liu2010indexability} and the random access methods. To ensure the fairness, we choose PPO to train the agent in one step DRL method. We call this method one step proximal policy optimization (OSPPO) method.
\subsection{Simulation Setting}
We simulate a shared spectrum consisting of $K=16$ orthogonal channels, where $12$ channels follow the fixed-pattern switching model and $4$ channels follow the correlated Markov model. The fixed-pattern switching subsets are divided into $4$ groups with $3$ channels in each group, where the channels in one group have a same state, and the switching probability between these groups $p_{\text{sw}}=0.8$. The correlated Markov subsets are divided into $2$ groups, where each group contains $2$ channels, and these $2$ groups follow the same transition probabilities $p_{01}=p_{10}=0.2$. The length of a time slot is $1$ ms. In each time slot, the CMCR transmits $N=16$ pulses with $T_r=50$ $\mu $s. In this paper, the interference to noise ratio (INR) is defined as $\frac{J_m}{N_0}$ and the signal to noise ratio (SNR) is defined as $\frac{\lvert \beta \rvert^2}{N_0\Delta f}$. In our simulation, the INR is set to $10$ dB. The hidden layer of the policy net and value net has $256$ units. In the training process, the learning rate is set to $1$e$-5$, $\gamma$ is set to $0.99$.

\subsection{Results and Discussions}
\begin{figure}[th]
    \centering
    \includegraphics[width=0.4\textwidth,trim={4pt 2pt 2pt 2pt},clip]{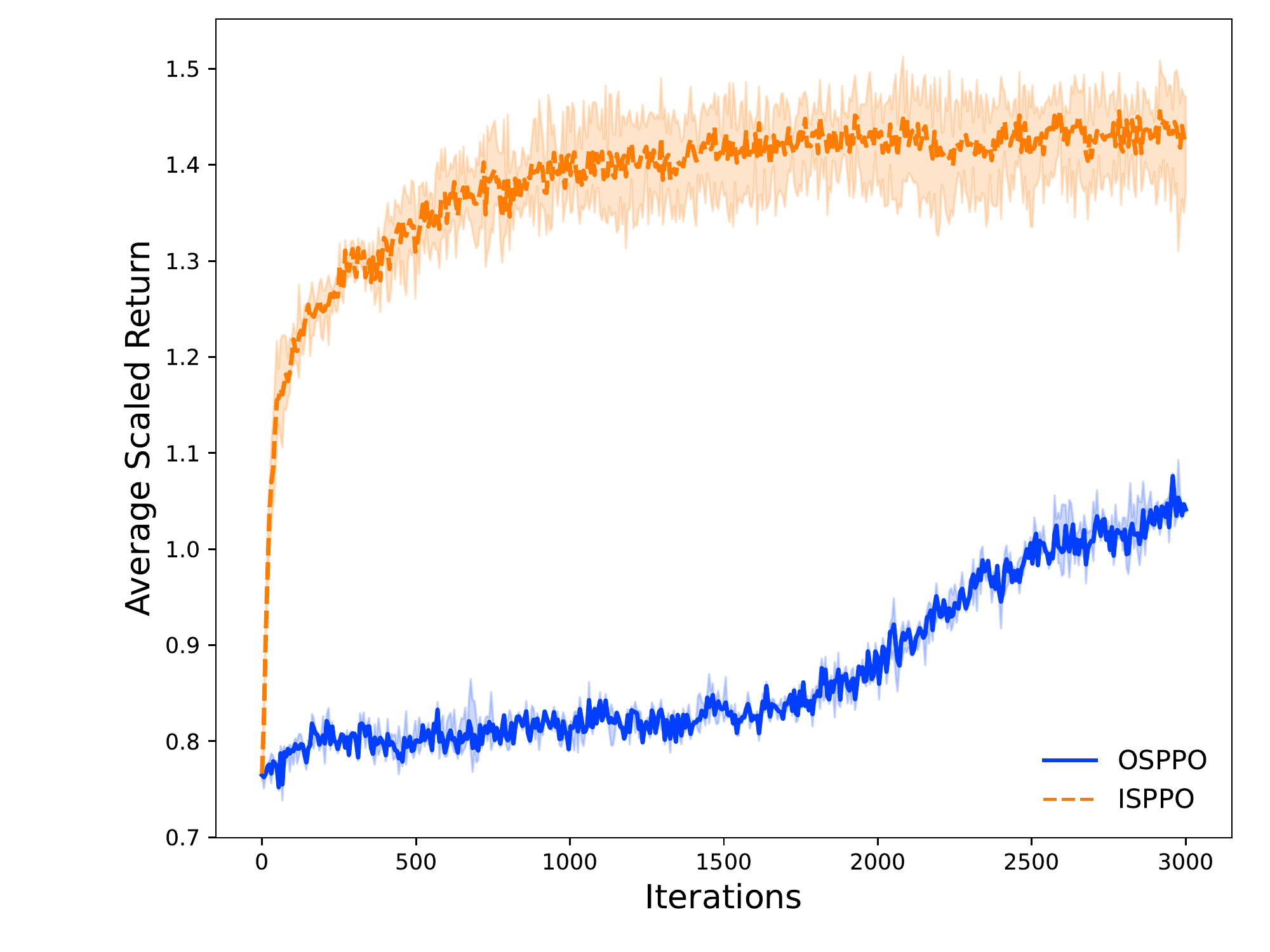}
    \caption{Learning curve of RL methods}
    \label{fig:training}
\end{figure}

The training performance of our proposed method against the existing deep reinforcement learning methods is shown in Fig. \ref{fig:training}, where the y-axis is the scaled cumulative discounted reward. In this experiment, we set $M=4$. It is shown that the proposed method converges much faster and achieves much better final performance among different random seeds. In each iteration, the DRL agent interacts with the environment for $512$ time slots. We can see that the proposed method gets a near optimal performance after $250$ iterations which totally takes $128$ s. The OSPPO method converges much slower, which results from a low sample efficiency. This is because the ISPPO method shares network parameters between sub-actions, and the output dimension of the policy network is no more than $M$. However, in the OSPPO, the output layer of the policy network has $\binom{K}{M}$ units. Too much redundant parameters make the OSPPO hard to converge.
To study the sensitivity of our proposed method, we conduct further experiments under the different switching probability $p_{\text{sw}}$ and the transition probability $p_{01},p_{10}$. The SNR of the waveforms is set to $10$ dB. We run our proposed method and the competitors for $L=2000$ time slots. The mean and standard deviation of the detection probabilities (with percentage) are shown in Table \ref{tabel.P_sens}. Overall, our proposed method performs robustly for different probabilities. As we can see, the DRL based methods perform much better than the whittle index policy, which proves that the correlation between channels has a strong impact on the performance of the whittle index policy.
\begin{table}[t]
\centering
\fontsize{\mytablefontsize}{\mytablebaselineskip\baselineskip}
\selectfont
\caption{\normalsize{Sensitivity study under different probabilities}}
\label{tabel.P_sens}
\begin{threeparttable}[h]
\begin{tabular}{ccccc}
\toprule
Case & ISPPO & OSPPO & Whittle & Random\\
\hline{}
                C1 & $91.5\pm23.1$ & $82.0\pm35.2$ & $62.7\pm41.0$  & $66.0\pm35.2$ \\
                C2 & $90.8\pm24.8$ & $87.1\pm27.5$ & $66.2\pm38.5$  & $62.3\pm37.8$ \\
                C3 & $90.9\pm25.2$ & $75.9\pm32.3$ & $68.9\pm40.4$  & $67.1\pm35.2$ \\
                C4 & $91.2\pm25.3$ & $87.5\pm30.2$ & $63.7\pm41.1$  & $64.4\pm36.7$ \\
\bottomrule
\end{tabular}

\begin{tablenotes}
\fontsize{8pt}{1}
     \item\scriptsize{C1}:$p_{\text{sw}}\!=\!0.80,p_{01}\!=\!p_{10}\!=\!0.20$; \scriptsize{C2}:$p_{\text{sw}}\!=\!0.80,p_{01}\!=\!p_{10}\!=\!0.15$, 
     \item\scriptsize{C3}:$p_{\text{sw}}\!=\!0.85,p_{01}\!=\!p_{10}\!=\!0.20$; \scriptsize{C4}:$p_{\text{sw}}\!=\!0.85,p_{01}\!=\!p_{10}\!=\!0.15$.
\end{tablenotes}

\end{threeparttable}
\end{table}

\begin{figure}[th]
    \centering
    \includegraphics[width=0.4\textwidth,trim={4pt 2pt 2pt 2pt},clip]{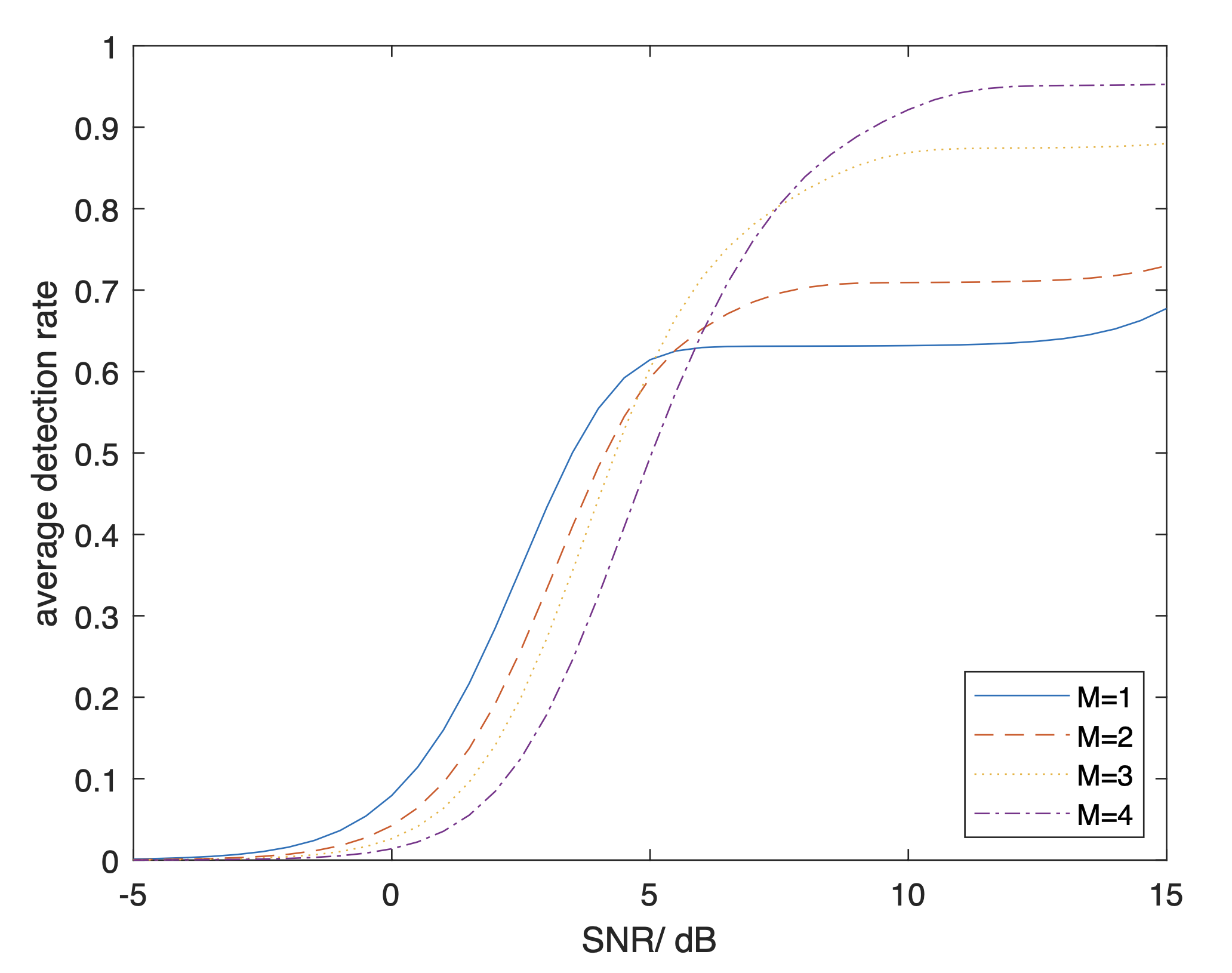}
    \caption{Detection rate under different frequency number}
    \label{fig:avg_det_rate}
\end{figure}

To show the influence of the carrier frequency number, we vary $M$ while keeping the total energy of the transmitted wave constant. Then, we evaluate the well trained ISPPO agent for $L=200$ time slots under the same environment. The false alarm rate is set to $1$e$-6$, and the average detection rate under the different SNR is shown in Fig. \ref{fig:avg_det_rate}. We can see that under the high SNR, the average detection rate increases with $M$. This is because the CMCR acquires more information about environment through more frequencies, and thus predicts the state of the environment more accurately. Meanwhile, the energy of the waveform is divided into several possible \textbf{good} frequencies, which reduces the risk that all of the waveform energy is interfered by the communication users. Under the low SNR, the CMCR suffers from a small detection loss because of the noncoherent integration between different frequencies. 

\section{Conclusion}
In this paper, we consider a problem where a CMCR shares the same spectrum resources with communication users and aims to minimize the interference from communication users. We formulated the problem into a MDP. To tackle the challenge of the combinatorial action space, we break a general MDP into a iteratively selecting MDP, where the action space is significantly reduced. To reduce the redundant parameters of the network, we share the value and policy network between different sub-actions. Moreover, the PPO algorithm is used to improve the training efficiency of the proposed method. Numerical experiments have been carried out to demonstrate the effectiveness of the proposed method. 
\section{References}
\bibliographystyle{ieeetr}
\bibliography{ref.bib}

\begin{thebibliography}{10}
\providecommand{\url}[1]{#1}
\csname url@samestyle\endcsname
\providecommand{\newblock}{\relax}
\providecommand{\bibinfo}[2]{#2}
\providecommand{\BIBentrySTDinterwordspacing}{\spaceskip=0pt\relax}
\providecommand{\BIBentryALTinterwordstretchfactor}{4}
\providecommand{\BIBentryALTinterwordspacing}{\spaceskip=\fontdimen2\font plus
\BIBentryALTinterwordstretchfactor\fontdimen3\font minus
  \fontdimen4\font\relax}
\providecommand{\BIBforeignlanguage}[2]{{%
\expandafter\ifx\csname l@#1\endcsname\relax
\typeout{** WARNING: IEEEtran.bst: No hyphenation pattern has been}%
\typeout{** loaded for the language `#1'. Using the pattern for}%
\typeout{** the default language instead.}%
\else
\language=\csname l@#1\endcsname
\fi
#2}}
\providecommand{\BIBdecl}{\relax}
\BIBdecl

\bibitem{9205659}
T.~Huang, N.~Shlezinger, X.~Xu \emph{et~al.}

\bibitem{1631800}
P.~Antonik, M.~Wicks, H.~Griffiths \emph{et~al.}, ``Frequency diverse array
  radars,'' in \emph{2006 IEEE Conference on Radar}, 2006, pp. 215--217.

\bibitem{8683655}
M.~Bică and V.~Koivunen, ``Multicarrier radar-communications waveform design
  for rf convergence and coexistence,'' in \emph{ICASSP 2019 - 2019 IEEE
  International Conference on Acoustics, Speech and Signal Processing
  (ICASSP)}, 2019, pp. 7780--7784.

\bibitem{8361480}
D.~Cohen, D.~Cohen, Y.~C. Eldar \emph{et~al.}, ``Summer: Sub-nyquist mimo
  radar,'' \emph{IEEE Transactions on Signal Processing}, vol.~66, no.~16, pp.
  4315--4330, 2018.

\bibitem{greco2018cognitive}
M.~S. Greco, F.~Gini, P.~Stinco \emph{et~al.}, ``Cognitive radars: On the road
  to reality: Progress thus far and possibilities for the future,'' \emph{IEEE
  Signal Processing Magazine}, vol.~35, no.~4, pp. 112--125, 2018.

\bibitem{sheng2020sensing}
X.~Sheng and S.~Wang, ``Sensing-transmission tradeoff for multimedia
  transmission in cognitive radio networks,'' in \emph{GLOBECOM 2020-2020 IEEE
  Global Communications Conference}.\hskip 1em plus 0.5em minus 0.4em\relax
  IEEE, 2020, pp. 1--6.

\bibitem{zhong2019deep}
C.~Zhong, Z.~Lu, M.~C. Gursoy \emph{et~al.}, ``A deep actor-critic
  reinforcement learning framework for dynamic multichannel access,''
  \emph{IEEE Transactions on Cognitive Communications and Networking}, vol.~5,
  no.~4, pp. 1125--1139, 2019.

\bibitem{9527128}
C.~E. Thornton, R.~M. Buehrer, and A.~F. Martone, ``Constrained contextual
  bandit learning for adaptive radar waveform selection,'' \emph{IEEE
  Transactions on Aerospace and Electronic Systems}, vol.~58, no.~2, pp.
  1133--1148, 2022.

\bibitem{liu2010indexability}
K.~Liu and Q.~Zhao, ``Indexability of restless bandit problems and optimality
  of whittle index for dynamic multichannel access,'' \emph{IEEE Transactions
  on Information Theory}, vol.~56, no.~11, pp. 5547--5567, 2010.

\bibitem{song2019solving}
H.~Song, H.~Jang, H.~H. Tran \emph{et~al.}, ``Solving continual combinatorial
  selection via deep reinforcement learning,'' \emph{arXiv preprint
  arXiv:1909.03638}, 2019.

\bibitem{schulman2017proximal}
J.~Schulman, F.~Wolski, P.~Dhariwal \emph{et~al.}, ``Proximal policy
  optimization algorithms,'' \emph{arXiv preprint arXiv:1707.06347}, 2017.

\bibitem{richards2014fundamentals}
M.~A. Richards, \emph{Fundamentals of radar signal processing}.\hskip 1em plus
  0.5em minus 0.4em\relax McGraw-Hill Education, 2014.

\bibitem{9354050}
L.~Wang, T.~Huang, Y.~Liu \emph{et~al.}, ``Randomized stepped frequency radars
  exploiting block sparsity of extended targets: A theoretical analysis,''
  \emph{IEEE Transactions on Signal Processing}, vol.~69, pp. 1378--1393, 2021.

\bibitem{wang2019network}
J.~Wang, S.~Guan, C.~Jiang \emph{et~al.}, ``Network association in
  machine-learning aided cognitive radar and communication co-design,''
  \emph{IEEE Journal on Selected Areas in Communications}, vol.~37, no.~10, pp.
  2322--2336, 2019.

\bibitem{li2021robust}
K.~Li, B.~Jiu, H.~Liu \emph{et~al.}, ``Robust antijamming strategy design for
  frequency-agile radar against main lobe jamming,'' \emph{Remote Sensing},
  vol.~13, no.~15, p. 3043, 2021.

\bibitem{kay1993fundamentals}
S.~M. Kay, \emph{Fundamentals of statistical signal processing}.\hskip 1em plus
  0.5em minus 0.4em\relax Prentice-Hall, Inc., 1993.

\bibitem{wang2018deep}
S.~Wang, H.~Liu, P.~H. Gomes \emph{et~al.}, ``Deep reinforcement learning for
  dynamic multichannel access in wireless networks,'' \emph{IEEE Transactions
  on Cognitive Communications and Networking}, vol.~4, no.~2, pp. 257--265,
  2018.

\bibitem{xu2020application}
Y.~Xu, J.~Yu, and R.~M. Buehrer, ``The application of deep reinforcement
  learning to distributed spectrum access in dynamic heterogeneous environments
  with partial observations,'' \emph{IEEE Transactions on Wireless
  Communications}, vol.~19, no.~7, pp. 4494--4506, 2020.

\bibitem{sutton1998introduction}
R.~S. Sutton, A.~G. Barto \emph{et~al.}, ``Introduction to reinforcement
  learning,'' 1998.

\bibitem{schulman2015high}
J.~Schulman, P.~Moritz, S.~Levine \emph{et~al.}, ``High-dimensional continuous
  control using generalized advantage estimation,'' \emph{arXiv preprint
  arXiv:1506.02438}, 2015.

\bibitem{mnih2016asynchronous}
V.~Mnih, A.~P. Badia, M.~Mirza \emph{et~al.}, ``Asynchronous methods for deep
  reinforcement learning,'' in \emph{International conference on machine
  learning}.\hskip 1em plus 0.5em minus 0.4em\relax PMLR, 2016, pp. 1928--1937.

\end{thebibliography}
\end{document}